\documentclass[twocolumn,showpacs,showkeys,preprintnumbers,amsmath,amssymb]{revtex4-1}
\usepackage{bbm}
\usepackage{amsfonts}

\usepackage{epsfig}
\usepackage{graphicx}
\usepackage{amssymb}   
\usepackage{dcolumn}
\usepackage{bm}
\usepackage[colorlinks,citecolor=blue,linkcolor=red,urlcolor=blue,hyperindex]{hyperref}
\usepackage{url}
\begin{document}

\vspace{2.5cm}
\title{Classical-driving-assisted quantum synchronization in non-Markovian environments}
\author{Xing Xiao$^{1}$}
\author{Tian-Xiang Lu$^{1}$}
\author{Wo-Jun Zhong$^{1}$}
\author{Yan-Ling Li$^{2}$}
\altaffiliation{liyanling0423@gmail.com}
\affiliation{$^{1}$ College of Physics and Electronic Information, Gannan Normal University, Ganzhou, China\\
$^{2}$School of Information Engineering, Jiangxi University of Science and Technology, Ganzhou 341000, China}

\begin{abstract}
We study the quantum phase synchronization of a driven two-level system (TLS) coupled to a structured environment and demonstrate that quantum synchronization can be enhanced by the classical driving field. We use the Husimi $Q$-function to characterize the phase preference and find the in-phase and anti-phase locking phenomenon in the phase diagram. Remarkably, we show that the in-phase classical driving enables a TLS to reach stable anti-phase locking in the Markovian regime. However, we find that the synergistic action of classical driving and non-Markovian effects significantly enhances the initial in-phase locking. By introducing the $S$-function and its maximal value to quantify the strength of synchronization and sketch the synchronization regions, we observe the typical signatures of the hollowed Arnold tongue in the parameter regions of synchronization. In the hollowed Arnold tongue, the synchronization regions exist both inside and outside the tongue while unsynchronized regions only lie on the boundary line. We also provide an intuitive interpretation of the above results by using the quasimode theory. 
\end{abstract}

\maketitle

\section{Introduction}
\label{intro}
Because of the direct or indirect interactions, the components of a system may adjust their own local dynamics to a common rhythm. This phenomenon is well known as synchronization~\cite{Blekhman1988}. The original study about synchronization could be dated back to 1657 when Christian Huygens found ``the sympathy of two clocks'', i.e., two pendulums always swung at the same frequency albeit in opposite directions to each other~\cite{Pena2016}. Examples of synchronization are widespread in multidisciplinary studies and in particular in physics, such as the coupled oscillators~\cite{Strogatz2000}. At present, the study of synchronization behavior in various kinds of real complex systems has become one of the active topics in different disciplines~\cite{Arenas2008}.

Unlike the ubiquity of synchronization in the classical world, the study of synchronization in the quantum regime has garnered increasing interest only recently~\cite{Orth2010,Lohe2010,Giorgi2013,Walter2015,Ameri2015,Fiderer2016,Giorgi2016,Witthaut2017,Bellomo2017,Militello2017,Sonar2018,Roulet2018a,Roulet2018b,Koppen2019,Giorgi2019,Karpat2019,Henriet2019,Eneriz2019,Karpat2020,Parra2020,Laskar2020,Kato2021,Cabot2021,Karpat2021,Zhou2021,Solanki2022,Zhang2022,Ali2022,Krithika2022,Li2022,Wachtler2022}.
Important advances have been made toward understanding quantum synchronization by
considering the quantum version of classically synchronous systems, for instance, Van der Pol oscillators \cite{Lee2013,Walter2014}.  Following this, the extension of synchronization to genuinely quantum systems without classical counterparts has been considered in atomic system \cite{Xu2014}, spin system \cite{Roulet2018b}, Bose–Einstein condensates \cite{Zhang2020}, and superconducting circuit system \cite{Quijandria2013}. The studies of quantum synchronization generally fall into two categories: spontaneous synchronization (or mutual synchronization) and forced synchronization. 
In the case of spontaneous synchronization, the interested system becomes synchronized in the transient evolution of dynamical systems due to the interaction between the subsystems or an external environment. Therefore, environmental noises play a significant role in enabling spontaneous synchronization in open quantum systems. Surprisingly, noise-induced synchronization has been reported in Refs.~\cite{Nakao2007,Imai2022,Schmolke2022}. This phenomenon highlights the constructive role of environmental noises in synchronization. In contrast to spontaneous synchronization, forced synchronization usually emerges with 
externally driven forces. This was considered in Refs.~\cite{Zhirov2008,Zhirov2009}, in which the behavior of one and two superconducting qubits coupled to a driven dissipative oscillator is demonstrated, and in Refs.~\cite{Roulet2018a,Parra2020}, where the forced phase synchronization in low-dimensional quantum systems is investigated. It is worth noting that whether a two-level system (TLS) could be synchronized has experienced a theoretical debate. The main disagreement is whether there is a valid limit cycle in TLS \cite{Roulet2018a}. In Ref.~\cite{Parra2020}, the authors predicted that such a valid limit cycle indeed exists in the stationary mixed state of the TLS, because each of those possible pure states that make up the mixed state would precess around the $z$-axis in the Bloch sphere. This theoretical prediction was subsequently confirmed by experiments in a trapped-ion system \cite{Zhang2022}.

Any realistic quantum system inevitably interacts with external environments. Therefore, the influence of environmental noises on forced synchronization is an intriguing problem that should be evaluated.
Recently, non-Markovian environments have drawn particular attention in quantum science and technology since the relevant environment's correlation time is not too small compared to the system’s relaxation time in many physical systems, in particular, artificial synthesis of materials \cite{Lambropoulos2000,Sakamoto2017,Carmele2019}. In fact, the valuable research of non-Markovian environments is the existence of information backflow from the environment (named as non-Markovian effects) \cite{Vega2017,Breuer2016,Gholipour2020,Tserkis2022,Poulsen2022}. The influences of non-Markovian effects on the dynamics of entanglement, quantum discord, and quantum Fisher information have been extensively explored in the past decades \cite{Bellomo2007,Xiao2009,Liu2011,Tong2010,Fanchini2010,Lu2010,Li2015}. Remarkably, the relationship between the degree of non-Markovianity and the onset of spontaneous synchronization of the qubit pair has been established by Karpat et al \cite{Karpat2021}. In a recent paper~\cite{Ali2022}, the authors discuss the synchronization of a single qubit embedded in a non-Markovian environment, and phase locking can be found outside the Arnold tongue region, which is in contrast to the previous studies where the synchronization happens within the tongue region~\cite{Roulet2018a,Parra2020}. Note that the discussions in \cite{Ali2022} are limited to considering the influence of
non-Markovian effects on spontaneous synchronization. However, what has been lacking thus far is a systematic study of the classical driving that could bring about quantum synchronization in non-Markovian environments. 
We believe this is an interesting problem since it gathers the center ingredients of both forced synchronization and
spontaneous synchronization.

Motivated by the above consideration, we will analyze a simple model consisting of a driven TLS coupled to a zero-temperature environment with Lorentzian spectral density, in order to discuss the role of classical driving in quantum synchronization and study the mechanism of in-phase synchronization in the non-Markovian regime.
To this aim, we use the Husimi $Q$-function to characterize the quantum synchronization phenomenon. We will show that there is anti-phase synchronization in Markovian regime in the presence of in-phase classical driving (The phrase ``in-phase'' means the phase difference between the TLS and the classical driving is close to 0). Intriguingly, the results in non-Markovian environments are different. We find that the phenomenon of in-phase locking occurs in long time dynamics with the assistance of classical driving. Two factors would be responsible for the in-phase synchronization in non-Markovian environments: the classical driving and the non-Markovian effects. The non-Markovian effects provide the feasibility of in-phase synchronization while the classical driving provides a feasible way to enhance it. We systematically investigate the parameter regions of synchronization and observe characteristic signatures of the hollowed Arnold tongue in the non-Markovian regime, which indicates that the synchronization regions exist both inside and outside the tongue.
The physical mechanism behind it can be explained by the quasimode theory. We argue that our model could be naturally generalized to the case of two driven TLS interacting with two independent non-Markovian environments.

This paper is organized as follows. In Sec.~\ref{sec:2}, we introduce the model and give the analytic solutions under the rotating-wave approximation. In Sec.~\ref{sec:3}, we investigate the features of synchronization of a driven TLS by calculating the Husimi $Q$-function. We show that long time, robust, in-phase synchronous dynamics occurs in the non-Markovian regime with the assistance of classical driving. In Sec.~\ref{sec:4}, we display how the synchronization region is affected by other factors, such as the Rabi frequency, the strength of non-Markovian effects, and the detuning. In Sec.~\ref{sec:5}, we show that the underlying mechanism of classical-driving-enhanced in-phase synchronization could be interpreted by the quasimode theory. Finally, we give a brief discussion and summarize our main results in Sec.~\ref{sec:6} .

\section{Description of the Model}
\label{sec:2} 

The model that we considered in this paper describes a TLS coupled to a zero-temperature bosonic environment . The TLS with frequency $\omega_0$ is driven by a classical single-mode ﬁeld $\mathcal{E}(t)=\mathcal{E}_{0}\cos(\omega_{L}t)$. Note that we have assumed the phase of the classical field is $\phi_{L}=0$. The Hamiltonian of system is described by ($\hbar=1$)
\begin{equation}
\label{eq1}
H_{S}=\frac{\omega_{0}}{2}\sigma_{z}+\frac{\Omega}{2}\left(\sigma_{+}e^{-i\omega_{L}t}+\sigma_{-}e^{i\omega_{L}t}\right).
\end{equation}
The Rabi frequency is $\Omega=-d_{eg}\mathcal{E}_{0}$ with $d_{eg}$ the transition dipole moment. $|e\rangle$ and $|g\rangle$ are the excited and ground states of the TLS. $\sigma_{j}(j=x,y,z)$ are the Pauli operators and $\sigma_{\pm}=(\sigma_{x}\pm i\sigma_{y})/2$. Considering the interaction between system and environment, the total system is described by the following Hamiltonian
\begin{equation}
\label{eq2}
H=H_{S}+\sum_{k}\omega_{k}b_{k}^{\dagger}b_{k}+\sum_{k}\left(g_{k}\sigma_{+}b_{k}+g_{k}^{*}b_{k}^{\dagger}\sigma_{-}\right),
\end{equation}
where $b_{k}^{\dagger}$ and $b_{k}$ are the creation and annihilation operators of the $k$th mode of the environment with frequency $\omega_{k}$. The coupling strength between the TLS and the $k$th mode is $g_{k}$. 

Using a unitary transformation $U_{r}=e^{-i\omega_{L}\sigma_{z}t/2}$, we can transfer the
Hamiltonian into the rotating reference frame. The effective Hamiltonian can be written as
\begin{equation}
\label{eq3}
H_{\rm eff}=\frac{\Delta}{2}\sigma_{z}+\frac{\Omega}{2}\sigma_{x}+\sum_{k}\omega_{k}b_{k}^{\dagger}b_{k}+\sum_{k}\left(g_{k}e^{i\omega_{L}t}\sigma_{+}b_{k}+\rm{H.c.}\right),
\end{equation}
where $\Delta=\omega_{0}-\omega_{L}$ denotes the detuning between the TLS frequency $\omega_0$ and classical driving field frequency $\omega_{L}$. In the dressed-state bases, the effective Hamiltonian will be expressed as
\begin{equation}
\label{eq4}
H_{\rm eff}=\frac{\omega_{D}}{2}\rho_{z}+\sum_{k}\omega_{k}b_{k}^{\dagger}b_{k}+\cos^2\frac{\eta}{2}\sum_{k}(g_{k}e^{i\omega_{L}t}\rho_{+}b_{k}+\rm{H.c.}),
\end{equation}
with the dressed frequency $\omega_{D}=\sqrt{\Delta^2+\Omega^2}$. The new operators $\rho_{z}=|E\rangle\langle E|-|G\rangle\langle G|$ and $\rho_{+}=|E\rangle\langle G|$ are defined by the dressed states 
$|E\rangle=\cos\frac{\eta}{2}|e\rangle+\sin\frac{\eta}{2}|g\rangle, |G\rangle=-\sin\frac{\eta}{2}|e\rangle+\cos\frac{\eta}{2}|g\rangle$ with $\eta=\tan^{-1}(\Omega/\Delta)$. 

We assume that the environment is initially in the vacuum state, then there is no more than one excitation in the total system. The Hilbert space will be restricted to the following subspace spanned in the dressed bases: 
\begin{eqnarray}
\label{eq5}
|\varphi_{0}\rangle&=&|G\rangle_{S}\otimes|\tilde{0}\rangle_{E},\nonumber\\
|\varphi_{1}\rangle&=&|E\rangle_{S}\otimes|\tilde{0}\rangle_{E},\\
|\varphi_{k}\rangle&=&|G\rangle_{S}\otimes|\tilde{1}_{k}\rangle_{E},\nonumber
\end{eqnarray}
where $|\tilde{0}\rangle_{E}=\bigotimes_{k=1}^{N}|0_{k}\rangle_{E}$ denotes the vacuum state, and $|\tilde{1}_{k}\rangle_{E}=\bigotimes_{j=1,j\neq k}^{N}|0_{j}\rangle_{E}|1_{k}\rangle_{E}$ denotes that there is one excitation in the
\emph{k}th mode of the environment. Notice that $|\varphi_{0}\rangle$ is immune to the environment, only the evolution of $|\varphi_{1}\rangle$ is dominated by Eq. (\ref{eq4}). The dynamical evolution can be mapped as
\begin{equation}
\label{eq6}
|\varphi_{1}\rangle\rightarrow q(t)|\varphi_{1}\rangle+\sum_{k}q_{k}(t)|\varphi_{k}\rangle,
\end{equation}
with the normalization $|q(t)|^2+\sum_{k}|q_{k}(t)|^2=1$.

Solving the Schr\"{o}dinger equation, we have
\begin{equation}
\label{eq7}
\dot{q}(t)=-\cos^4\frac{\eta}{2}\int_{0}^{t}d\tau f(t-\tau)q(\tau),
\end{equation}
where the integral kernel $f(t-\tau)$ can be expressed in terms of the spectral density function $J(\omega)$ of the environment as  \cite{Breuer2002}.
\begin{equation}
\label{eq8}
f(t-\tau)=\int d\omega J(\omega
)\exp\Big[i(\omega_{D} +\omega_{L}-\omega)(t-\tau)\Big].
\end{equation}

In the above, we have assumed the environment has a continuum of frequencies such that they can be characterized by the spectral density function $J(\omega)$. Clearly, the solution of Eq.~(\ref{eq8}) highly depends on the explicit form of $J(\omega)$. We focus on the Lorentzian spectral density of the form
\begin{equation}
\label{eq9}
 J(\omega)=\frac{\gamma_{0}\lambda^{2}}{2\pi}\frac{1}{(\omega_{0}-\omega-\delta)^{2}+\lambda^{2}},
\end{equation}
where $\delta=\omega_0-\omega_{c}$ is the detuning between the TLS frequency $\omega_0$ and
the center frequency of the Lorentzian spectrum $\omega_{c}$. 
$\lambda$ is the spectral width which is related to the environmental correlation time $\tau_{\rm E}=\lambda^{-1}$, while 
$\gamma_0$ is the coupling strength between the TLS and the environment which is connected to the decay rate of the
TLS. According to the relationship between the parameters $\lambda$ and $\gamma_0$, we can define the weak and strong coupling regimes. 
In the weak-coupling regime ($\gamma_0<\lambda/2$), the dynamics of the TLS is Markovian. On the contrast, the dynamics of the TLS will be non-Markovian in the strong-coupling regime ($\gamma_0>\lambda/2$).

Substituting Eq.~(\ref{eq9}) into Eq.~(\ref{eq8}), we can obtain the analytical form of correlation function $f(t-\tau)$ by Laplace transform
\begin{equation}
\label{eq10}
f(t-\tau)=\frac{\gamma_{0}\lambda}{2}\exp\left[-K(t-\tau)\right],
\end{equation}
with $K=\lambda+i\Delta-i\delta-i\omega_{D}$.
Then the probability amplitude $q(t)$ can be expressed as
\begin{equation}
\label{eq11}
q(t)=e^{-\frac{Kt}{2}}\Big[\cosh(\frac{\Gamma t}{4})+\frac{2K}{\Gamma}\sinh(\frac{\Gamma t}{4})\Big],
\end{equation}
with $\Gamma=\sqrt{4K^2-2\gamma_{0}\lambda(1+\cos\eta)^2}$. 

Therefore, the evolution map of the TLS in the dressed bases will be 
\begin{eqnarray}
\label{eq12}
&&|\varphi_{0}\rangle\rightarrow|\varphi_{0}\rangle,\\
&&|\varphi_{1}\rangle\rightarrow q(t)|\varphi_{1}\rangle+\sqrt{1-|q(t)|^2}\bm{|\varphi_{k}\rangle},\nonumber
\end{eqnarray}
where $\bm{|\varphi_{k}\rangle}=\big[\sum_{k=1}^{N}q_{k}(t)|\tilde{1}_{k}\rangle_{E}/\sqrt{1-|q(t)|^2}\big]$.

\begin{figure*}
  \includegraphics[width=1\textwidth]{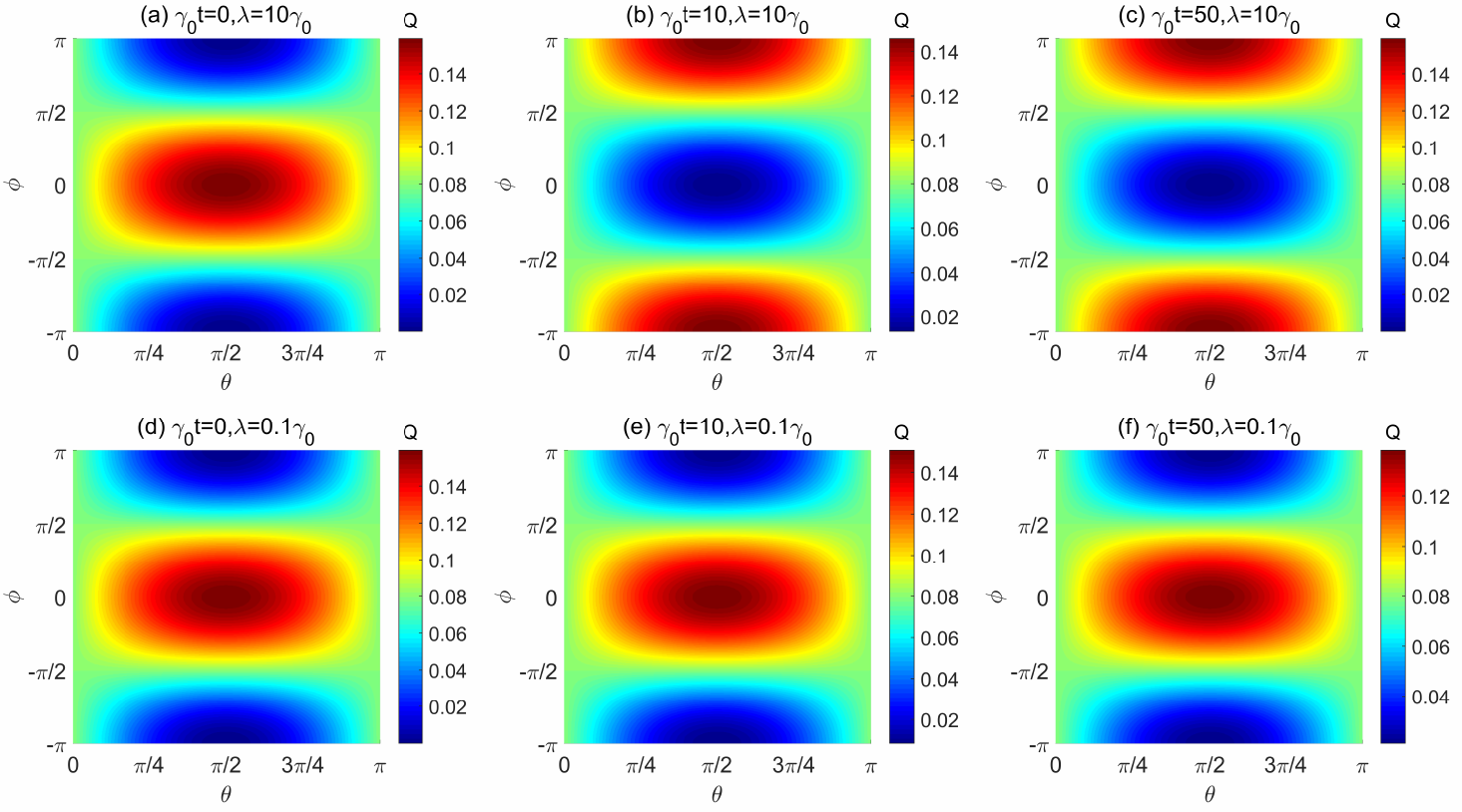}
\caption{(color online) The transient synchronization characterized by Husimi $Q$-function under the classical driving. (a)-(c) In the Markovian regime with $\lambda=10\gamma_{0}$. (d)-(f) In the non-Markovian
regime with $\lambda=0.1\gamma_{0}$.
The other parameters are $\Omega=2\gamma_0$, $\Delta=0$ and $\delta=0$. }
\label{fig1}       
\end{figure*}

\section{Classical-driving-assisted Quantum Synchronization}
\label{sec:3}
With above state map in mind, the dynamics of the TLS for an arbitrary initial state 
can be obtained by tracing the environment's
degrees of freedom
\begin{eqnarray}
\rho(0)=\left(\begin{array}{ll}
\rho_{00}(0) & \rho_{01}(0) \\
\rho_{10}(0) & \rho_{11}(0)
\end{array}\right)\rightarrow\rho(t)=\left(\begin{array}{ll}
\rho_{00}(t) & \rho_{01}(t) \\
\rho_{10}(t) & \rho_{11}(t)
\end{array}\right)\nonumber\\
\label{eq13}
\end{eqnarray}
in the bases of $\{|e\rangle,|g\rangle\}$. The elements have the explicit expressions

\begin{eqnarray}
\rho_{00}(t)&=& \rho_{00}(0)\big[\cos^{2}\frac{\eta}{2}-\cos \eta \sin^{2}\frac{\eta}{2}|q(t)|^{2}+\mathcal{A}\big]\nonumber\\
&+&\rho_{01}(0)\big[-\frac{1}{4}\sin 2\eta|q(t)|^{2} +\mathcal{B}\big]\nonumber\\
&+&\rho_{11}(0)\big[\cos^{2}\frac{\eta}{2}-\cos \eta \cos^{2}\frac{\eta}{2}|q(t)|^{2}-\mathcal{A}\big]\nonumber\\
&+&\rho_{10}(0)\big[-\frac{1}{4}\sin 2\eta|q(t)|^{2} +\mathcal{B}^{*}\big],\\
\label{eq14}
\rho_{01}(t)&=& \rho_{00}(0)\big[ -\frac{1}{2}\sin\eta+\sin \eta \sin^{2}\frac{\eta}{2}|q(t)|^{2}+\mathcal{B}\big]\nonumber\\
&+&\rho_{01}(0)\big[\frac{1}{2}\sin^{2}\eta|q(t)|^{2} +\cos^{4}\frac{\eta}{2}q^{*}(t)+\sin^{4}\frac{\eta}{2}q(t)\big]\nonumber\\
&+&\rho_{11}(0)\big[-\frac{1}{2}\sin\eta+\sin \eta \cos^{2}\frac{\eta}{2}|q(t)|^{2}-\mathcal{B}\big]\nonumber\\
&+&\rho_{10}(0)\big[\frac{1}{2}\sin^{2}\eta|q(t)|^{2}-\mathcal{A}\big],
\label{eq15}
\end{eqnarray}
with $\rho_{11}(t)=1-\rho_{00}(t)$, $\rho_{10}(t)=\rho_{01}^{*}(t)$ and $\mathcal{A}=\frac{1}{4}\sin^{2}\eta[q(t)+q^{*}(t)]$, $\mathcal{B}=\frac{1}{2}\sin\eta[\cos^{2}\frac{\eta}{2}q^{*}(t)-\sin^{2}\frac{\eta}{2}q(t)]$.

In order to discuss synchronization, we employ the Husimi $Q$-function as a measure to evaluate the phase preference of the TLS. Husimi $Q$-function is one of the simplest distributions of quasiprobability in phase space, which is defined for any TLS state $\rho$ as
\begin{equation}
Q(\theta,\phi)=\frac{1}{2\pi}\langle \theta,\phi|\rho|\theta,\phi\rangle,
\label{eq16}
\end{equation}
where $|\theta,\phi\rangle=\cos\frac{\theta}{2}|1\rangle+\sin\frac{\theta}{2}\exp(i\phi)|0\rangle$ are the eigenstates of the operator $\hat{\vec{\sigma}}\cdot\hat{\vec{n}}$ with $\vec{n}=(\sin\theta\cos\phi,\sin\theta\sin\phi,\cos\theta)$. The Husimi $Q$-function provides a phase portrait, which allows us to visualize any state $\rho$ in terms of the coherent states $|\theta,\phi\rangle$. According to Eqs.~(\ref{eq13})-(\ref{eq15}), the explicit form of the Husimi $Q$-function can be written as
\begin{eqnarray}
Q(\theta,\phi,t)&=&\frac{1}{2\pi}\Big[\rho_{11}(t)\cos^{2}\frac{\theta}{2}+\frac{1}{2}\rho_{10}(t)\sin\theta e^{i\phi} \nonumber\\
&+&\frac{1}{2}\rho_{01}(t)\sin\theta e^{-i\phi}+\rho_{00}(t)\sin^{2}\frac{\theta}{2}\Big]. 
\label{eq17}
\end{eqnarray}

In the following text, we will discuss the phase synchronization of the driven TLS assuming that the initial state is
\begin{equation}
|\psi(0)\rangle_{S}=\frac{1}{\sqrt{2}}\big(|g\rangle+|e\rangle\big).
\label{eq18}
\end{equation}

\begin{figure}
  \includegraphics[width=0.5\textwidth]{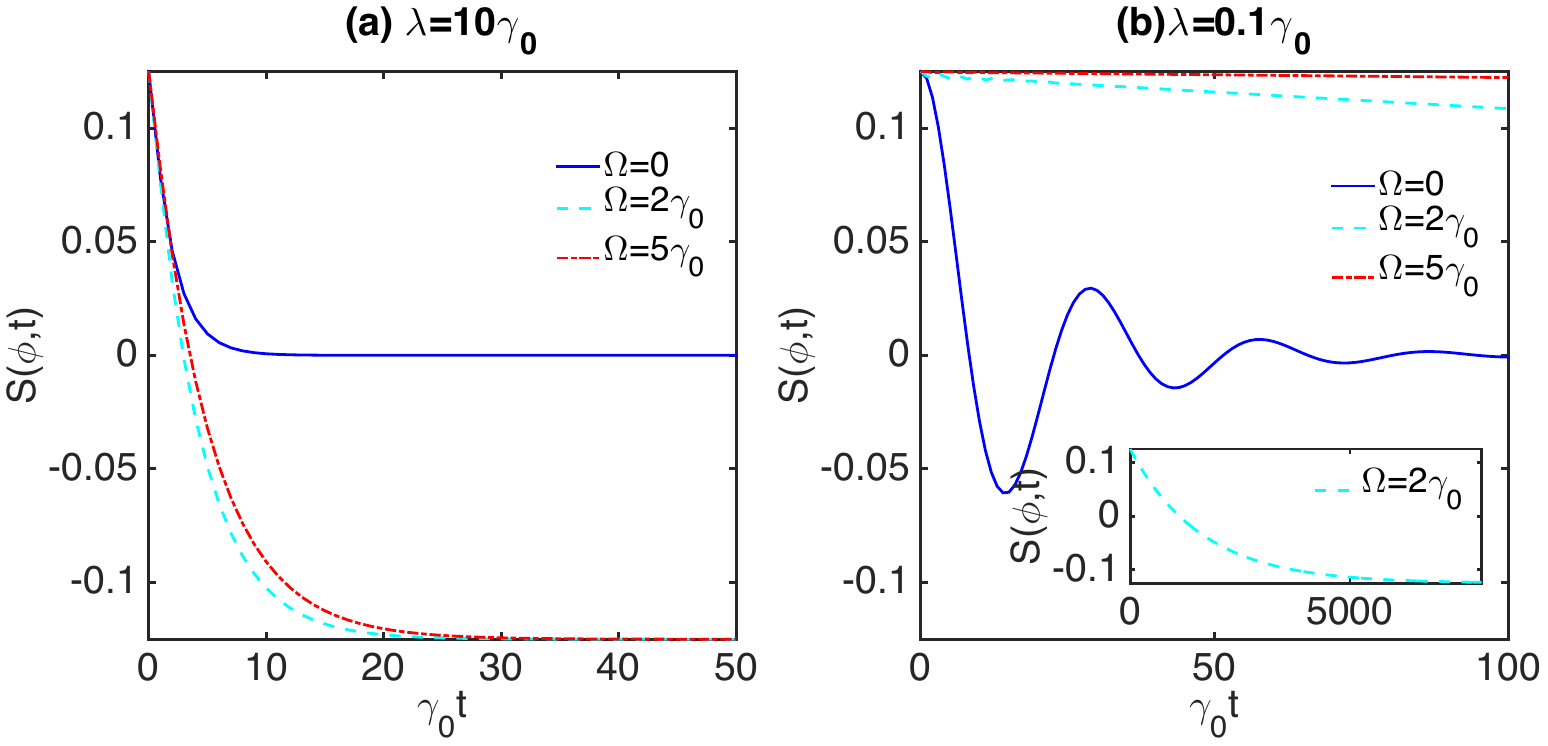}
\caption{(color online) The synchronization measure $S(\phi)$ as a
function of $\gamma_{0}t$, (a) in the
Markovian regime with $\lambda=10\gamma_{0}$, (b) in the non-Markovian
regime with $\lambda=0.1\gamma_{0}$. The other
parameters are $\phi=0$, $\Delta=0$ and $\delta=0$. }
\label{fig2}       
\end{figure}

As a preliminary exploration, in Fig. \ref{fig1}, we first show the transient synchronization characterized by Husimi $Q$-function in the presence of classical driving. We compare the results in both non-Markovian and Markovian environments, and show how in-phase locking preserves in one case but doesn't occur in the other case. In Figs.~\ref{fig1}(a)-(c), the transient dynamics of $Q$-function are shown in the Markovian regime. It is clearly shown that the initial phase preference peaked at $\phi=0$ with $\theta=\pi/2$ rapidly turns to be dipped at the same point. Similarly, the initially dipped points $(\phi=\pm\pi,\theta=\pi/2)$ change to peaked points. This is a phenomenon of anti-phase locking since the phase difference between the TLS and classical driving field is close to $\pi$. Notice that our result is significantly different from that obtained in Ref.~\cite{Ali2022}, where the initial phase preference decreases and is eventually wiped out as time passes. Accordingly, the distribution of $Q$-function eventually becomes uniform along the $\phi$-axis in their paper~\cite{Ali2022}. The only difference is that we have considered a \emph{driven} TLS. As we will see in Fig.~\ref{fig2}(a), the phenomenon of anti-phase locking is present with the classical driving and absent without the classical driving. Figures~\ref{fig1}(d)-(f) show the transient dynamics of $Q$-function in the non-Markovian regime. In contrast to the Markovian case, the $Q$-function shows a dynamical in-phase locking. The initial phase preference preserves for a long time with the assistance of classical driving. 

In order to observe the effect of classical driving on phase locking clearly, we would like to describe the phase preference not only qualitatively but also quantitatively. Following the works done in Refs.~\cite{Roulet2018a,Parra2020}, we introduce a synchronization measure $S(\phi)$ by integrating over the angular variable $\theta$. 
\begin{eqnarray}
S(\phi)&=&\int_{0}^{\pi}d\theta\sin\theta Q(\theta,\phi)-\frac{1}{2\pi}\\
&=&\frac{\Re\big[\rho_{01}(t)\big]\cos\phi+\Im\big[\rho_{01}(t)\big]\sin\phi}{4}. \nonumber
\label{eq19}
\end{eqnarray}
Notice that $S(\phi)$ is zero if and only if there is no phase synchronization between the TLS and the classical driving field, i.e., the uniform distribution of $Q$-function. On the other hand, a nonzero value of $S(\phi)$ implies the existence of phase locking. Specifically, a positive value of $S(\phi)$ indicates the in-phase locking while a negative of this measure means the anti-phase locking. It is natural to expect that the sign of $S(\phi)$ would change over varied $\phi$, but the corresponding relationship between in-phase locking (anti-phase locking) and $S>0$ ($S<0$) doesn’t change because the in-phase or anti-phase relation between the TLS and the classical driving is defined by the phase difference of them. It is in-phase locking when the phase difference is nearly 0 while anti-phase locking when the phase difference is close to $\pi$. .

Figure~\ref{fig2} would be useful for understanding the synchronization dynamics exhibited in Fig.~\ref{fig1}. We show the dynamics of $S(\phi,t)$ with respect to different strengths of classical driving. The first conclusion is that indeed, in both
Markovian and non-Markovian regimes, no phase locking is expected when there is no classical driving (i.e., the blue solid lines in Fig.~\ref{fig2}). 
We can observe the rapid decay of $S(\phi,t)$ in the Markovian regime in Fig.~\ref{fig2}(a). In the absence of classical driving, it turns to be zero, which is in accordance with the conclusion in Ref.~\cite{Ali2022}. The reason is that it doesn't exist a stable limit cycle in this case. The stationary state of the TLS is $|g\rangle$, which remains at the south pole of the Bloch sphere. Such a state can not provide a valid limit cycle for synchronization because it is lying exactly on the rotation axis \cite{Roulet2018a}. However, when the classical driving is performed on the TLS, the coexistence of driving and damping ensures a stationary state which, in the Bloch representation, lies on the meridian of the $x-z$ plane except for the south and north poles. Thus, the limit cycle is constructed under the driving since the stationary state can precess around the $z$-axis once we move back to the non-rotating frame \cite{Parra2020}.
We observe that the phase preference undergoes an asymptotical change from in-phase locking to anti-phase locking under the classical driving in Fig.~\ref{fig2}(a). This also can be understood from the steady-state solution corresponding to the initial state of Eq.~(\ref{eq18}). In the long time limit, the state of TLS reduces to $|\psi(t\rightarrow\infty)\rangle=\cos\frac{\eta}{2}|g\rangle-\sin\frac{\eta}{2}|e\rangle$. Thus, the phase difference between the TLS and the classical diving field is shifted from 0 (in-phase locking) to $\pi$ (anti-phase locking), which also can be confirmed by Figs.~\ref{fig1}(a)-(c).

Figure~\ref{fig2}(b) shows the dynamics of $S(\phi,t)$ in the non-Markovian regime. In the absence of classical driving, the synchronization measure $S(\phi,t)$ vanishes after a few damped oscillations induced by the non-Markovian effects (i.e., the information backflow). It is interesting to note that classical driving can dramatically enhance the in-phase locking in the non-Markovian regime. The larger the Rabi frequency is, the better the in-phase locking. The underlying reason is that the presence of classical driving will shift the resonant frequency of the TLS, which reduces the information exchange between the TLS and the environment. Then the outflow of the information from the TLS is suppressed. Particularly, such suppression is much more striking in the non-Markovian regime, which ensures the in-phase locking phenomenon. 
This means that the phenomenon of in-phase locking shouldn't be attributed solely to either classical driving or non-Markovian effects, but the combination of them. The non-Markovian effects offer the feasibility of in-phase locking while the classical driving provides a feasible way to enhance it. The inset of Fig.~\ref{fig2}(b) indicates that it is also anti-phase locking in the non-Markovian regime in the long time limit because the stationary state of the driven TLS is the same in both Markovian and non-Markovian baths.

\section{Synchronization Regions and the Hollowed Arnold Tongue}
\label{sec:4}
In order to deeply understand the synchronization, we have to study how the synchronization regions are determined by the other parameters, such as the coupling strength, the spectral width, the Rabi frequency, and the detuning $\delta=\omega_0-\omega_{c}$. Considering $S(\phi)$ depends on the value of $\phi$, which only captures the phase preference in a specific orientation. The maximum of $S(\phi)$ varying $\phi$ from $-\pi$ to $\pi$, would be better to represent the synchronization strength. 

\begin{figure*}
  \includegraphics[width=1\textwidth]{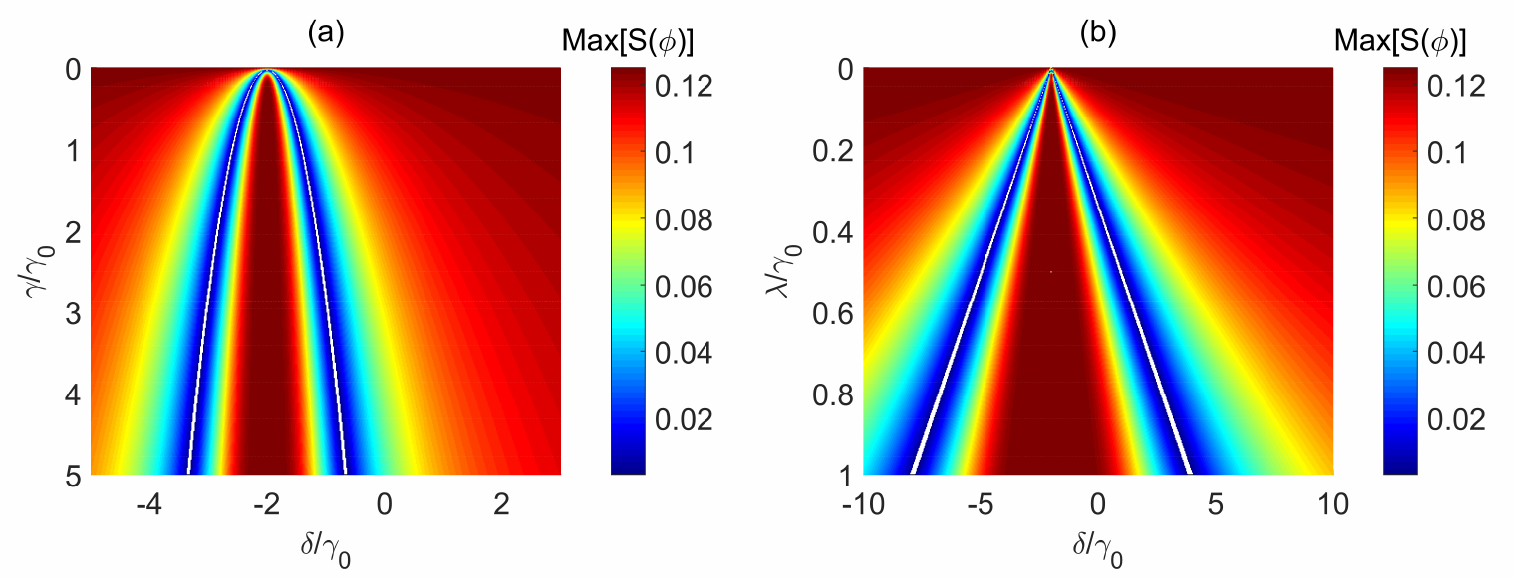}
\caption{(color online) The hollowed Arnold tongue. (a) The maximum of $S(\phi)$ as function of the detuning $\delta=\omega_0-\omega_c$ and the coupling strength $\gamma$ (in units of $\gamma_{0}$) in the non-Markovian regime with $\lambda=0.1\gamma_0$. (b) The maximum of $S(\phi)$ as function of the detuning $\delta=\omega_0-\omega_c$ and the spectral width $\lambda$. The white line is the boundary line of the hollowed Arnold tongue.
The other parameters are $\gamma_0 t=100$, $\Omega=2\gamma_0$, $\Delta=0$. }
\label{fig3}       
\end{figure*}

In Fig.~\ref{fig3}(a), we show how the synchronization region is governed by the detuning between the TLS frequency $\omega_0$ and the spectral center frequency $\omega_{c}$ for different coupling strength $\gamma$ in the non-Markovian regime. To vary the coupling strength, we consider the coupling strength $\gamma$ in spectral density $J(\omega)$ varying in units of $\gamma_{0}$. The contour plot of maximal $S(\phi)$ is similar to the Arnold tonguelike phase diagrams \cite{Pikovsky2001}. It is worth mentioning that the presence of classical driving makes the synchronization regions different from those found in previous literatures, in which the synchronization region only exists either inside or outside the tongue. However, in our model, it is remarkable that the synchronization regions exist both inside and outside the tongue. We name this phase diagram as the \emph{hollowed} Arnold tongue. The hollowed Arnold tongue takes the white lines as the boundary line, on which there is no synchronization (i.e., $\rm{Max}[S(\phi)]=0$). In the outer area of the boundary line, the region of synchronization increases with the decrease of the $\gamma$. This is straightforward as the weak coupling between TLS and the environment favors enhancing the in-phase synchronization. However, the trend inside the tongue is exactly the opposite of the outside situation, which implies that the region of synchronization becomes larger with stronger coupling. The reason for this counterintuitive phenomenon is that we have traversed all possible values of $\phi\in[-\pi,\pi]$ to find the phase preference regardless of whether it is in-phase synchronization or anti-phase synchronization. It is in-phase synchronization on the outside of the tongue and anti-phase synchronization on the inside of the tongue. Namely, the locked phases inside and outside the tongue are different.
Another intriguing feature is that the central line of the tongue region is not at $\delta=0$, but at $\delta=\Delta-\sqrt{\Delta^2+\Omega^2}$, which is shifted by the classical driving field. This could be understood from the quasimode theory, as we will show in Sec.~\ref{sec:5}.

The maximum of $S(\phi)$ as function of the detuning $\delta$ and the spectral width $\lambda$ is shown in Fig.~\ref{fig3}(b). We observe a triangular-shaped Arnold tongue. In the external area of the tongue, it appears to indicate that decreasing the spectral width can improve the resulting in-phase synchronization. On the other hand, the increase of the spectral width will enlarge the anti-phase synchronization region inside the tongue. This antagonism, as we explained above, comes from the different phase preferences.

\begin{figure*}
 \includegraphics[width=1\textwidth]{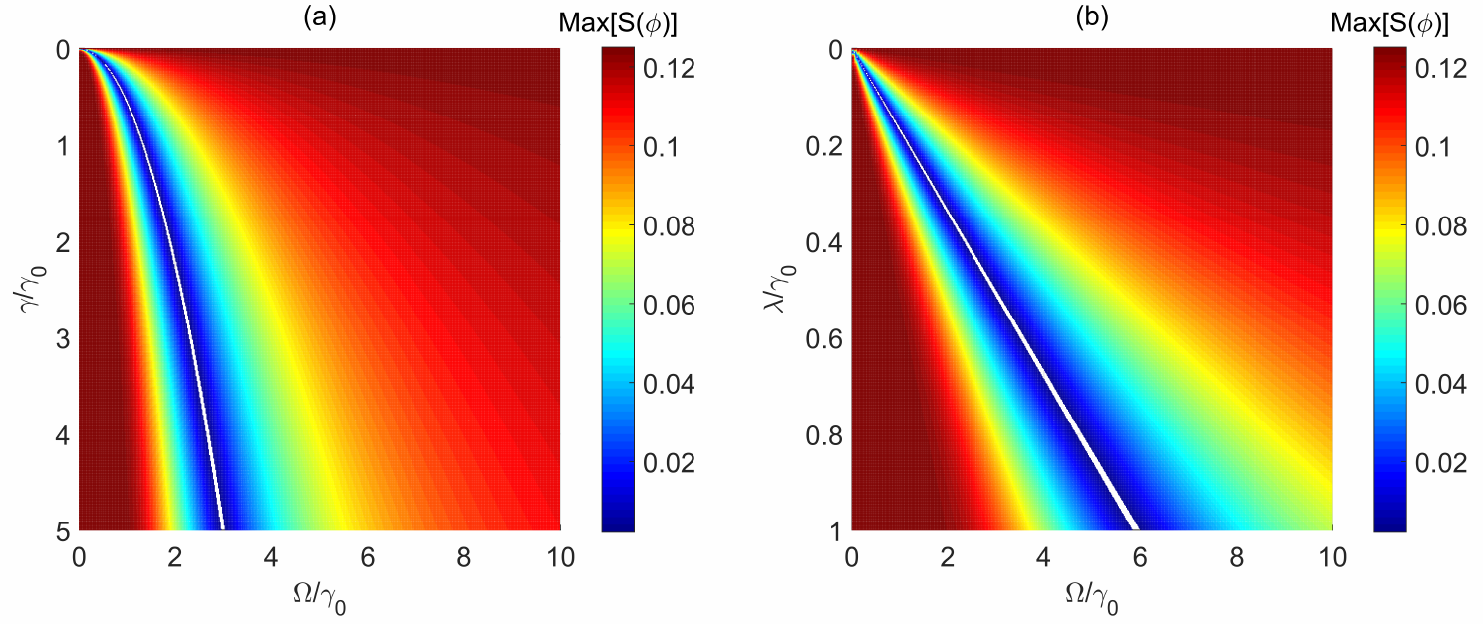}
\caption{(color online) The half Arnold tongue. (a) The maximum of $S(\phi)$ as function of the Rabi frequency $\Omega$ and the coupling strength $\gamma$ (in units of $\gamma_{0}$) in the non-Markovian regime with $\lambda=0.1\gamma_0$. (b) The maximum of $S(\phi)$ as function of the Rabi frequency $\Omega$ and the spectral width $\lambda$. The white line is the boundary line of the half Arnold tongue. The other parameters are $\gamma_0 t=100$, $\delta=0$, $\Delta=0$.}
\label{fig4}       
\end{figure*}

Figure~\ref{fig4}(a) is plotted to show how the contour plot of maximal $S(\phi)$ is determined by the parameters of Rabi frequency $\Omega$ and the coupling strength $\gamma$, while Fig.~\ref{fig4}(b) is determined by the Rabi frequency $\Omega$ and the spectral width $\lambda$. Notice that the synchronization region is in a half Arnold tongue distribution because the Rabi frequency is non-negative. Once again, we find the synchronization regions exist both inside and outside the half tongue. As we expect, the in-phase synchronization becomes more pronounced with the increase of Rabi frequency. These features can be nicely confirmed by the half Arnold tongue, shown in Figs.~\ref{fig4}(a) and (b) for the given values of $\gamma$ and $\lambda$. The increase of Rabi frequency will drive the synchronization region from anti-phase locking to in-phase locking. Although we have only shown the patterns of Arnold tongue at $\gamma_{0}t=100$ in Figs. \ref{fig3} and \ref{fig4}, the basic features of the hollowed Arnold tongue and half Arnold tongue always exist at different times except $\gamma_{0}t=0$ and $\gamma_{0}t=\infty$.

\section{Physical Interpretation}
\label{sec:5}

In this section, we try to understand the phenomenon of classical-driving-assisted quantum synchronization in non-Markovian environment in a more intuitive physical insight. For the Hamiltonian shown in Eq.~(\ref{eq4}), we can
move to a frame rotating with the frequency of the driving by applying a unitary transformation $U=\exp[i\omega_{L}\rho_{z}t/2]$.
\begin{equation}
H_{\rm eff}=\frac{\omega_{0}'}{2}\rho_{z}+\sum_{k}\omega_{k}b_{k}^{\dagger}b_{k}+\sum_{k}(g_{k}'\rho_{+}b_{k}+\rm{H.c.}),
\label{eq20}
\end{equation}
where $\omega_{0}'=\omega_0+\sqrt{\Delta^2+\Omega^2}-\Delta$ and $g_{k}'=\cos^{2}(\eta/2)g_{k}$. This effective Hamiltonian has the same form as the Hamiltonian of spontaneous decay of a TLS. The only difference is that Eq.~(\ref{eq20}) is written in the dressed state bases $\{|E\rangle,|G\rangle\}$. 

According to the quasimode theory \cite{Garraway1997,Dalton1999,Dalton2001}, there is only one discrete quasimode for the Lorentz spectral distribution of the environment. Therefore, the quasimode Hamiltonian can be divided into three parts
\begin{equation}
H_{\rm quasi}=H_{\rm SE}+H_{\rm memory}+H_{\rm dissipation},
\label{eq21}
\end{equation}
with
\begin{eqnarray}
\label{eq22}
&&H_{\rm SE}=\frac{1}{2}\omega_{0}'\rho_{z}+\omega _{c}D^{\dagger }D+\int\nu C^{\dagger}(\nu)C(\nu)d\nu,\nonumber\\
&&H_{\rm memory}=\sqrt{\frac{\gamma_{0}\lambda}{2}}\big(\rho_{+}D+\rho_{-}D^{\dagger}\big),\\
&&H_{\rm dissipation}=\sqrt{\frac{\lambda}{\pi}}\int d\nu\big[D^{\dagger}C(\nu)+DC^{\dagger}(\nu)\big],\nonumber
\end{eqnarray}
where $D$ and $C(\nu)$ are the annihilation operators of the discrete quasimode and the continuum quasimode of frequency $\nu$, respectively. $H_{\rm memory}$ denotes the coupling between the TLS and the discrete quasimode, while $H_{\rm dissipation}$ represents the interaction between the discrete and continuum quasimodes.

\begin{figure}
  \includegraphics[width=0.5\textwidth]{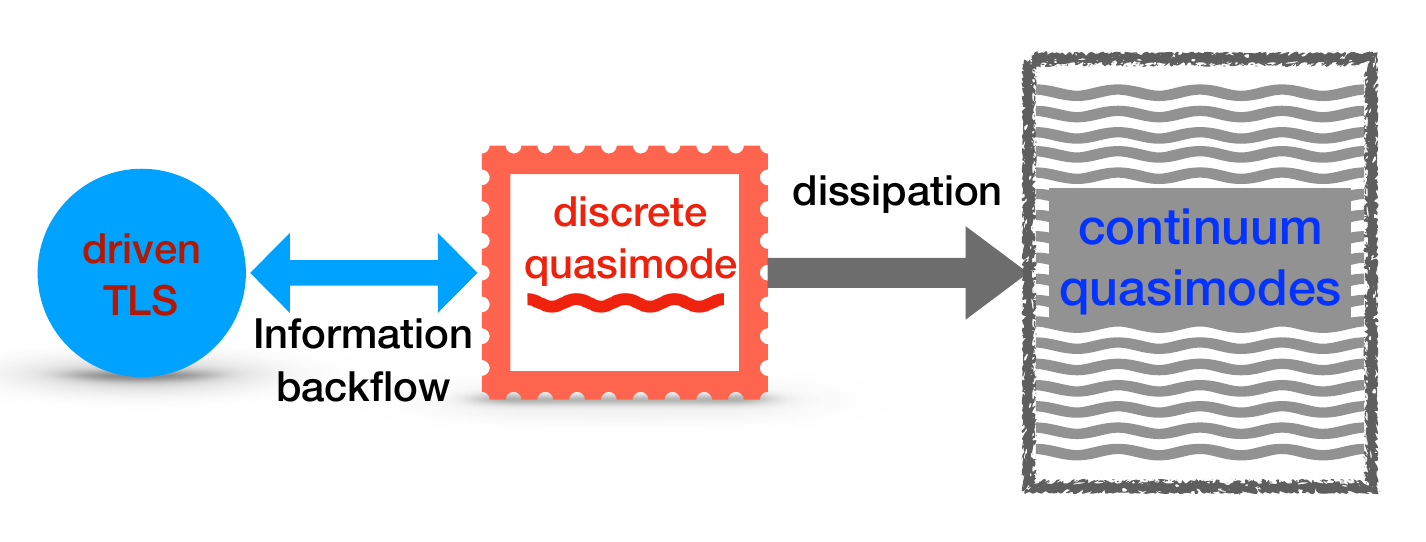}
\caption{(color online) The schematic diagram of quasimode picture of a driven TLS coupled to a non-Markvian environment. The TLS is coupled to the discrete quasimode, which is in turn coupled to the continuum quasimodes.}
\label{fig5}       
\end{figure}

The schematic diagram of quasimode picture is illustrated in Fig.~\ref{fig5}. The original modes of the non-Markovian environment have been transformed into two parts: the discrete and continuum quasimodes which behave as coupled quantum harmonic oscillators. The discrete mode can be understood as the resonant quasimode of a cavity, while the continuum modes are the external quasimodes outside the cavity. The discrete quasimode
functions as a memory between the TLS and the dissipative continuum quasimodes. The coupling between the TLS and the discrete quasimode could be depicted by the Jaynes-Cummings model. Dissipation only emerges from the interaction between the discrete and continuum quasimodes which is directly proportional to the spectral width $\lambda$.

Now we shall explore the physical mechanism of in-phase locking in non-Markovian environments. As we discussed in Fig.~\ref{fig2}(b), the combination of classical driving and non-Markovian effects are responsible for the enhancement of in-phase locking in non-Markovian environments. Figure~\ref{fig5} tells us that the discrete quasimode
functions as a memory. The information flows spontaneously from the TLS to discrete quasimode and in turn to continuum quasimodes. The presence of classical driving can increase the detuning between the TLS and discrete quasimode. The effective detuning between the TLS and discrete quasimode yields to
\begin{equation}
\Delta_{\rm eff}(\Omega, \Delta,
\delta)=\sqrt{\Delta^2+\Omega^2}-\Delta+\delta,
\label{eq23}
\end{equation}
which is proportional to the Rabi frequency $\Omega$. The effective resonant coupling occurs when $\Delta_{\rm eff}=0$ or $\delta=\Delta-\sqrt{\Delta^2+\Omega^2}$, which determines the central line of the tongue in Fig.~\ref{fig3}.

It is well known that the large effective detuning shall reduce the coupling between the TLS and the mode of the field \cite{Scully1997}. The result is the suppression of information flow from the TLS to the discrete quasimode. However, the presence of classical driving is only one-half of the story. As shown in Eq.~(\ref{eq20}), the dissipation is dominated by the spectral width $\lambda$. In the Markovian regime ($\lambda\gg\gamma_0$), the information stored in the memory (i.e., the discrete quasimode) decays exponentially into the 
continuum quasimodes. That is why classical driving is not valid for in-phase locking in the Markovian regime. On the contrary, in the non-Markovian regime $\lambda\ll\gamma_0$, the dissipative
interaction is weak and the information could be stored for a long time. It will exchange back
and forth between the TSL and discrete quasimode before it
completely flows into the continuum quasimodes. With the assistance of classical driving, the information could be well protected, as shown in Fig.~\ref{fig2}(b). Therefore, we can confirm the conclusion: the non-Markovian effects offer the feasibility of in-phase synchronization while classical driving provides a practical method to enhance it.

\section{Conclusions and Discussion}
\label{sec:6}

In summary, we have investigated the quantum phase synchronization of a driven
TLS coupled to a zero-temperature bosonic environment. In the Markovian regime, we have shown that the presence of a classical field will drive the phase preference to the $\phi=\pi$ direction, which is known as anti-phase locking. 
However, the initial phase preference could be locked for a quite long time
in the non-Markovian environment with the assistance of classical driving. Such a phenomenon of in-phase locking should be attributed to the synergistic action of classical driving and non-Markovian effects. To quantify the strength of synchronization and depict the synchronization regions, we have introduced the $S$-function and its maximal value as synchronization measures. We have systematically discussed how the synchronization regions are determined by different systemic parameters and observed the typical features of the Arnold tongue for a synchronized system. 
The remarkable result in our model is that the Arnold tongue is hollowed. The synchronization regions exist both inside and outside the tongue while the unsynchronized region marks out the boundary of the tongue. We point out that the locked phases inside and outside the tongue are different. Finally, we have provided an intuitive physical interpretation for the classical-driving-assisted quantum phase synchronization according to the quasimode theory.
These findings provide valuable tools for synchronizing in the open quantum systems via reservoir engineering.

In the previous discussion, we have restricted our analysis to a specific initial state (\ref{eq18}) of the TLS. Some discussion of other initial states seems necessary, such as $|\psi(0)\rangle_{S}=\alpha|g\rangle+\beta e^{i\phi_{s}}|e\rangle$, where $\phi_{s}$ is the phase of TLS and $\alpha^2+\beta^2=1$. In fact, the weight parameter $\alpha$ does not have any substantial effect on phase synchronization. We notice that the phase parameter $\phi_{s}$ can be divided into two categories: initially in-phase with the classical driving when $\phi_{s}\in(0,\pi/2)$ and initially anti-phase with the classical driving when $\phi_{s}\in(\pi/2,\pi]$. For the former, it is straightforward to conclude that the results are completely similar to the original, except that the values of $Q$ and $S$ change a little, but doesn't change in positive or negative signs. However, when $\phi_{s}\in(\pi/2,\pi]$, the results will be different because the initial phase relationship between the TLS and the classical driving has become anti-phase. The TLS will evolve directly from the initial state to the anti-phase steady state without going through a transition from in-phase locking to anti-phase locking. Accordingly, the synchronization measure $S$ will always be negative, which means that there is only anti-phase locking and no in-phase locking in both Markovian and non-Markovian baths. Thus our conclusions related to in-phase locking and hollowed Arnold tongue will be not valid in the latter case.

\begin{acknowledgements}
X. Xiao is supported by the National Natural Science Foundation of China under Grant Nos. 12265004 and 11805040. Y. L. Li is supported by Jiangxi Provincial Natural Science Foundation under Grant No. 20212ACB211004. T. X. Lu is supported by the National Natural Science Foundation of China under Grant No. 12205054.
\end{acknowledgements}



\end{document}